\documentclass[]{raa}            
\usepackage{graphicx,times}
\usepackage{natbib}

\providecommand{\bm}[1]{\mbox{\boldmath$#1$\unboldmath}}

\begin{document}

\title{Spectroscopic Study of Globular Clusters in the Halo of
  M31 with Xinglong 2.16m Telescope II: Dynamics, Metallicity and Age
 $^*$
\footnotetext{\small $*$ Supported by the National Natural Science Foundation of China.}
}

\volnopage{ {\bf 2011} Vol.\ {\bf 9} No. {\bf 00}, 000--000}
\setcounter{page}{1}

\author{Zhou Fan
  \inst{1}
  \and Ya-Fang Huang
  \inst{1,2}
  \and Jin-Zeng Li
  \inst{1}
  \and Xu Zhou
  \inst{1}
  \and Jun Ma
  \inst{1}
  \and Yong-Heng Zhao
  \inst{1}
}

\institute{Key Laboratory of Optical Astronomy, National Astronomical Observatories, Chinese Academy of Sciences, Beijing 100012, China; {\it zfan@bao.ac.cn}
  \and
  Graduate University of Chinese Academy of Sciences, Beijing 100049, China\\
  \vs \no
  {\small Received [year] [month] [day]; accepted [year] [month] [day] }
}

\abstract{ In our Paper I, we performed the spectroscopic observations 
    of 11 confirmed GCs in M31 with the Xinglong 2.16m telescope and we 
    mainly focus on the fits method and the metallicity gradient 
    for the M31 GC sample. In this paper, we analyzed and discussed more 
    about the dynamics, metallicity and age, and their distributions as well 
    as the relationships between these parameters. In our work, eight more 
    confirmed GCs in the halo of M31 were 
    observed, most of which lack the spectroscopic information before. 
    These star clusters are located far from the galactic 
    center at a projected radius of $\sim14$ to $\sim117$ kpc, which are more 
    spatially extended than that in the previous work. The Lick
    absorption-line indices and the radial velocities have been measured 
    primarily. Then the ages, metallicities $\rm [Fe/H]$ and $\rm [\alpha/Fe]$
    have been fitted by comparing the observed spectral feature indices and the 
    SSP model of Thomas et al. in the Cassisi and Padova stellar evolutionary 
    tracks, respectively. Our results show that most of the star clusters of 
    our sample are older than 10 Gyr except B290$\sim 5.5$ Gyr, and most of 
    them are metal-poor with the metallicity $\rm [Fe/H]<-1$, suggesting that
    these clusters were born at the early stage of the galaxy's formation. 
    We find that the metallicity gradient for the outer halo clusters with 
    $r_p>25$ kpc may not exist with a slope of $-0.005\pm0.005$ dex kpc$^{-1}$ 
    and if the outliers G001 and H11 are excluded, the slope dose not change 
    significantly with a value of $-0.002\pm0.003$ dex kpc$^{-1}$. We also 
    find that the metallicity is not a function of age for the GCs with 
    age $<7$ Gyr while for the old GCs with age $>7$ Gyr there seems to be a 
    trend that the older ones have lower metallicity. Besides, We plot 
    metallicity distributions with the largest sample of M31 GCs so far and 
    it shows the bimodality is not significant and the number of the metal-poor 
    and metal-rich groups becomes comparable. The spatial distributions shows 
    that the metal-rich group is more centrally concentrated while the 
    metal-poor group is occupy a more extended halo and the young population 
    is centrally concentrated while the old populaiton is more extended 
    spatially to the outer halo.
   }

\authorrunning{Fan et al.}            
\titlerunning{Spectroscopic Study of M31 Halo GCs}  
\maketitle


%
%
\section{Introduction}           
\label{intro.sec}

One way to better understand the formation and evolution of the
galaxies is through detailed studies of globular clusters (GCs),
which are often considered to be the fossils of galactic formation
and evolution processes, since they formed at the early stages
of their host galaxies' life cycles \citep{bh00}. GCs are densely
packed, very luminous, which usually contains several thousands to
approximately one million stars. Therefore, they can be detected
from great distances and are suitable as probes for studying the
properties of extragalactic systems. Since the halo globular
clusters (HGCs) are located far away from the galaxy center, they
are very important and useful to study the dark matter distribution
of the galaxy. Besides, since the HGCs are far from the galaxy
center, the background of galaxy becomes much lower, which makes the
observations much easier, compared to the disc GCs in the projected
direction of galaxies.

As the nearest ($\sim780$ kpc) and large spiral galaxy in our Local
Group, M31 (Andromeda) contains a great number of GCs from
$460\pm70$ \citep{bh01} to $\sim$530 \citep{per10}, which is an
ideal laboratory for us to study the nature of the HGCs. A great many
of new M31 HGCs have been discovered in the recent years,
which are important to study the formation history of M31 and its
dark matter content. \citet{h04} discovered nine previously unknown
HGCs of M31 using the INT survey. Subsequently,
\citet{h05} found three new, extended GCs in the halo of M31, which
have characteristics between typical GCs and dwarf galaxies.
\citet{mac06} reported four extended, low-surface-brightness
clusters in the halo of M31 based on {\sl Hubble Space
Telescope}/Advanced Camera for Surveys (ACS) imaging. These star
clusters are structurally very different from typical M31 GCs. On
the other side, since they are old and metal-poor, they look like
the typical Milky Way GCs. \citet{h07} found 40 new extended GCs in
the halo of M31 (out to $\sim100$ kpc from the galactic center)
based on INT and CFHT imaging. These extended star clusters in the
M31 halo are very similar to the diffuse star clusters (DSCs)
associated with early-type galaxies in the Virgo Cluster reported by
\citet{peng06} based on the ACS Virgo Cluster survey. Indeed, the
evidence shows that DSCs are usually fainter than typical GCs.
Later, \citet{mac07} reported 10 outer-halo GCs in M31, at $\sim$15
kpc to 100 kpc from the galactic center, eight of which were newly
discovered based on deep ACS imaging. The HGCs in their sample
are very luminous, compact with low metallicity, which are quite
different from their counterparts in our Galaxy. More recently,
\citet{ma10} constrained the age, metallicity, reddening and
distance modulus of B379, which also is an HGCs of M31, based on
the multicolor photometry.

In \citet{fan11} (hereafter Paper I) we observed 11 confirmed star
clusters, most of which are located in the halo of M31, with the OMR
spectrograph on 2.16m telescope at Xinglong site of National
Astronomical Observatories, Chinese Academy of Sciences, in fall of
2010. We estimated the ages, metallicities, $\alpha$-elements with
the SSP models as well as the the radial velocities and they found
that most of the halo clusters are old and metal-poor, which were
supposed to be born at the early stage of the galaxy formation
history. In this paper, we will continue the study of the HGCs of
M31 with the same instruments and a larger sample. This allows us to
be able to better understand the properties of the M31 outer halo.
This paper is organized as follows. In \S \ref{sam.sec} we describe
how we selected our sample of M31 GCs and their spatial
distribution. In \S \ref{obs.sec}, we reported the spectroscopic
observations with 2.16 m telescope and how the data was reduced and
the radial velocities and Lick indices were measured and calibrated.
Subsequently, in \S \ref{fit.sec}, we derive the ages, metallicities
and $\alpha$-element with $\chi^2-$minimization fitting. We also
discuss our final results on the metallicity distribution in the M31
halo. Finally, we summarized our work and give our conclusions in \S
\ref{sum.sec}.

\section{Sample selection}
\label{sam.sec}

The sources were selected from the updated Revised Bologna Catalogue
of M31 globular clusters and candidates \citep[RBC v.4, available
from http://www.bo.astro.it/M31;][]{gall04,gall06,gall07,gall09},
which is the latest and most comprehensive M31 GC catalogue so far.
The catalogue contains 2045 objects, including 663 confirmed star
clusters, 604 cluster candidates, and 778 other objects that were
previously thought to be GCs but later proved to be stars,
asterisms, galaxies, or H{\sc ii} regions. Indeed, many of the halo
clusters were from \citet{mac07}, who reported 10 GCs in the outer
halo of M31 from their deep ACS images, of which eight were detected
for the first time (see for details in \S \ref{intro.sec}). In our
work, our sample clusters are completely selected from RBC v.4. We
selected the confirmed and luminous clusters as well as being
located as far as they could from the galaxy center, where the local
background is too luminous for our observations. Finally, there are
eight bright confirmed clusters in our sample, all of which are
located in the halo of the galaxy. These clusters lack spectroscopic
observational data, especially for the metallicity measurements.
Thus it is necessary to observe the spectra of our sample clusters
systematically and constrain the spectroscopic metallicities and
ages in detail.

The observational information of our sample GCs are listed in
Table~\ref{t1.tab}, which includes the names, coordinates, projected
radii in kpc, exposures and observation dates. All the coordinates
(R.A. and Dec. in Cols. 2 and 3) and projected radii from the galaxy
center $r_{\rm p}$ (Col. 4) are all from RBC v.4, which were
calculated with M31 center coordinate $00:42:44.31$, $+41:16:09.4$
\citep{per02}, $PA=38^{\circ}$ and distance $d=785$ kpc
\citep{mcc05}.

\begin{table}[ht!!!]
\small
\centering
\begin{minipage}[]{100mm}
\caption[]{The observations of our sample GCs.}\label{t1.tab}\end{minipage}
\tabcolsep 3mm
\begin{tabular}{lccccc}
  \hline\noalign{\smallskip}
  ID & R.A. & Dec. & $r_p$&Exposure& Date \\
      & (J2000) & (J2000) & (kpc) &(second)&  \\
  \hline\noalign{\smallskip}
  B289    & 00:34:20.882 & +41:47:51.14 & 22.65 & 6000 &  08/28/2011 \\
  B290    & 00:34:20.947 & +41:28:18.18 & 21.69 & 7200 &  09/01/2011 \\
  H11     & 00:37:28.028 & +44:11:26.41 & 42.10 & 5400 &  09/01/2011 \\
  H18     & 00:43:36.030 & +44:58:59.30 & 50.87 & 5400 &  08/29/2011 \\
  SK108A  & 00:47:14.240 & +40:38:12.30 & 14.47 & 3600 &  08/28/2011 \\
  SK112A  & 00:48:15.870 & +41:23:31.20 & 14.28 & 5400 &  08/29/2011 \\
  MGC1    & 00:50:42.459 & +32:54:58.78 & 117.05& 3600 &  08/28/2011 \\
  H25     & 00:59:34.560 & +44:05:39.10 & 57.35 & 5400 &  09/01/2011 \\
  \noalign{\smallskip}\hline
\end{tabular}
\end{table}

\begin{figure}
\resizebox{\hsize}{!}{\rotatebox{0}{\includegraphics{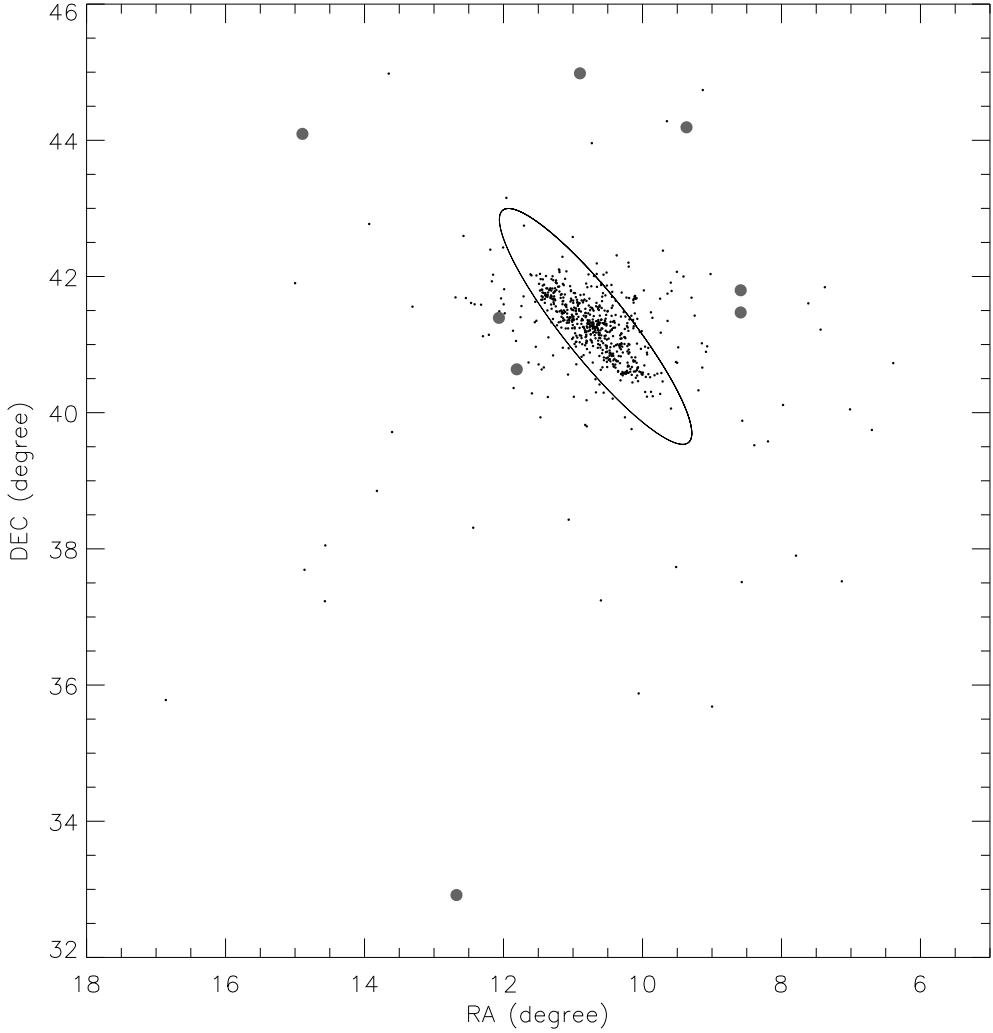}}}
\caption{Spatial distribution of M31 GCs. Our sample halo GCs are shown 
  with filled circles and the confirmed GCs from RBC v.4 are marked 
  with points. The large ellipse is the M31 disk/halo boundary as defined 
  in \citet{rac91}.}
\label{fig1}
\end{figure}

We show the spatial distribution of our sample eight halo GCs and all the
confirmed GCs from RBC v.4 in Figure~\ref{fig1}. The large ellipse is
the M31 disk/halo boundary as defined by \citet{rac91}.
Note that all of our sample are located in the halo of M31, which can
help us to access the nature of galaxy halo with an enlarged cluster
sample, compared to \citet{fan11}.

\section{Observations and data reduction}
\label{obs.sec}

Our Low-resolution spectroscopic observations were all taken at
the 2.16m optical telescope at Xinglong Site, which belongs to
National Astronomical Observatories, Chinese Academy of Sciences (NAOC),
from August 28th to September 1st, 2011 (Please see
Table~\ref{t1.tab}). An OMR (Optomechanics Research Inc.)
spectrograph and a PI 1340${\times}$400 CCD detector were used during
this run with a dispersion of 200 {\AA} mm$^{-1}$, 4.8 {\AA} pixel$^{-1}$,
and a 3.0 \arcsec slit. The typical seeing there was $\sim2.5$ \arcsec.
The spectra cover the wavelength range of $3500-8100$ {\AA} at 4 {\AA}
resolution. All our spectra have $S/N \ge 40$.

In order to calibrate our 2.16m data onto the Lick system, we also observed
eleven Lick standard stars (HR 6806, HR 6815, HR 7030, HR 7148, HR 7171,
HR 7503, HR 7504, HR 7576, HR 7977, HR 8020, HR 8165) near our field, which
are selected from a catalogue of all 25 index measurements and coordinates
for 460 stars \citep[ref, available from
http://astro.wsu.edu/worthey/html/system.html;][]{wo97,w94a}.
Most of these standard stars are luminous ($\sim5-6$ in V band), hence
the exposure time we took was 20 second with the OMR system.

The spectroscopic data were reduced following the standard procedures with
NOAO Image Reduction and Analysis Facility (IRAF v.2.15)
software package. First, the spectra have been bias and flat-field
corrected, as well as cosmic-ray removed. Then the wavelength calibrations were
performed based on Helium/Argon lamps exposed at both the beginning
and the end of the observations in each night. Flux calibrations were
performed based on observations of at least two of the
KPNO spectral standard stars \citep{mass88} each night. The atmospheric
extinction was corrected with the mean extinction coefficients
measurements of Xinglong through the Beijing-Arizona-Taiwan-Connecticut (BATC)
multicolor sky survey (H. J. Yan 1995, priv. comm.).

Before the Lick indices were measured, the heliocentric radial
velocities $V_r$ were measured by comparing the absorption
lines of our spectra with the templates in various radial
velocities. The typical internal velocity errors on a single measure
is $\sim 20$ km s$^{-1}$. The estimated radial velocities $V_r$ with
the associated uncertainties (Col. 2) are listed in
Table~\ref{t2.tab}. The published radial velocities $V_r$ (Col. 3)
are also listed for comparisons. The systematic difference between our
observed velocity and the catalogue velocity is found to be $\rm
29\pm39~km~s^{-1}$ and the standard deviation of the differences
between our observed velocity and the catalogue velocity is $\rm 78~km~
s^{-1}$ for the five pairs of the radial velocities. It suggests that our 
measurements agree with those listed in RBC v.4 since the systematic
difference between our measurements and the published values is not
significant. 

\begin{figure}
  \resizebox{\hsize}{!}{\rotatebox{0}{\includegraphics{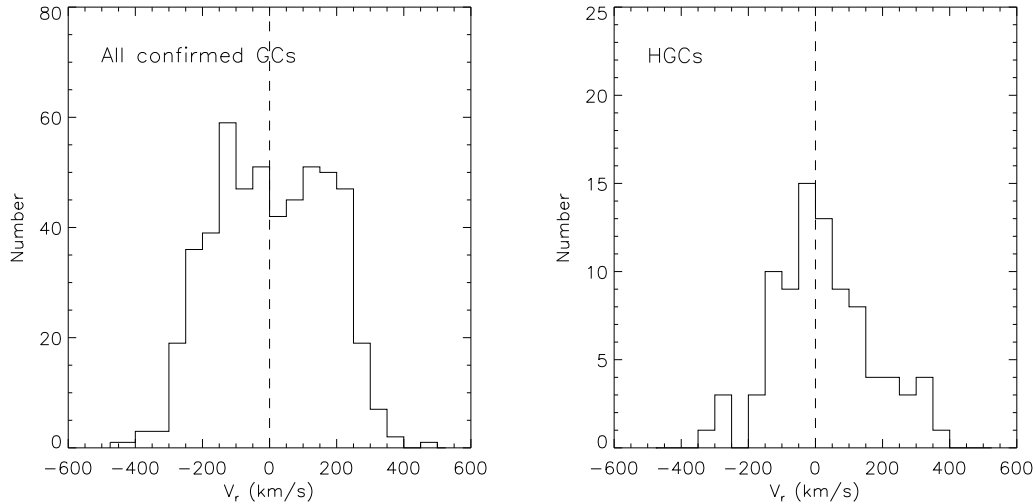}}}
  \caption{The distributions of radial velocity $V_r$ (corrected for the 
    systemic velocity of M31). {\it Left}: all the confirmed GCs. {\it
      Right}: the HGCs only.}
  \label{fig2}
\end{figure}

Figure~\ref{fig2} shows the radial velocity $V_r$ (corrected for
the systemic velocity of M31) as a function of the projected radii
from the galaxy center. The {\it Left} panel is for the all confirmed GCs
which have the radial velocity $V_r$ measurements and the {\it Right}
panel is for the HGCs, which refers to the GCs in the galaxy halo defined 
in Figure~\ref{fig1}. It can be noted that the radial velocity
distributions are basically symmetric in distributions 
either for all the confirmed GC sample or for the HGCs only.

\begin{table}[ht!!!]
\small
\centering
\begin{minipage}[]{100mm}
  \caption[]{The radial velocities $V_r$ of our sample GCs as well as
    the previous results.}\label{t2.tab}\end{minipage}
\tabcolsep 3mm
\begin{tabular}{lcc}
  \hline\noalign{\smallskip}
  ID & our work & RBC v.4 \\
  \hline\noalign{\smallskip}
B289    &  $-96.81\pm47.27$ & $-181\pm30$ \\
B290    & $-488.73\pm43.14$ & $-381\pm26$ \\
H11     & $-173.02\pm39.63$ &             \\
H18     & $-300.48\pm79.65$ &             \\
SK108A  & $-352.17\pm19.18$ & $-379\pm38$ \\
SK112A  & $-342.68\pm32.81$ & $-252\pm46$ \\
MGC1    & $-412.67\pm17.13$ & $-355\pm2$  \\
H25       & $-256.49\pm55.28$ &             \\
  \noalign{\smallskip}\hline
\end{tabular}
\end{table}

\begin{figure}
  \resizebox{\hsize}{!}{\rotatebox{0}{\includegraphics{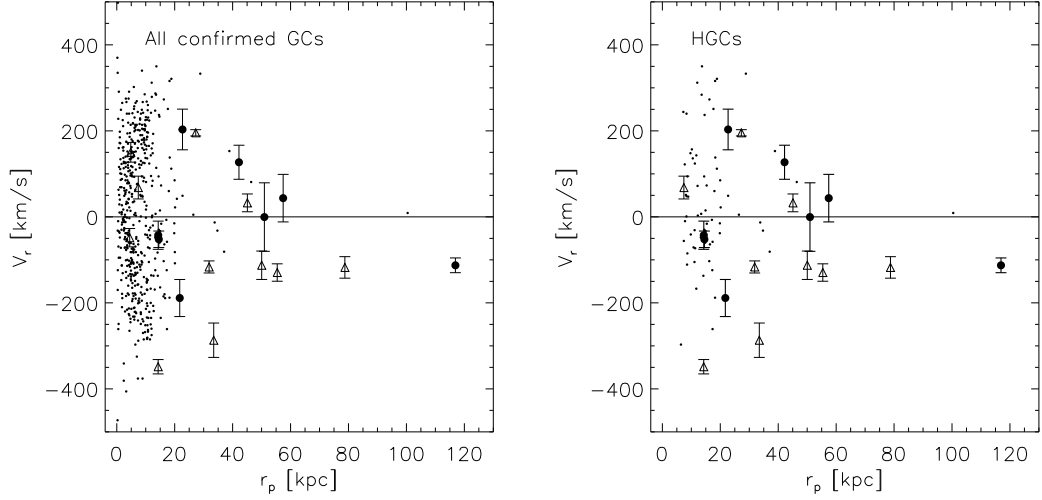}}}
  \caption{The radial velocity $V_r$ (corrected for the systemic
    velocity of M31) as a function of the projected radius. {\it Left}:
    all confirmed clusters and {\it Right}: the halo clusters.
    The filled circles with errors are the halo GCs from our sample
    while the points represent the velocities from RBC v.4 catalogue.}
  \label{fig3}
\end{figure}

We plotted the radial velocities $V_r$ versus the projected radii $r_p$
in Figure~\ref{fig3} where the radial velocities have been corrected for
the systemic velocity of M31 galaxy of $300\pm4$ km
s$^{-1}$\citep{per02}. The {\it Left} panel is for all the confirmed 
clusters in RBC v.4 while the right panel is for the halo
clusters which are defined in Figure~\ref{fig1}. The points are
the published measurements from RBC v.4 while the open triangles and
the filled circles with errors are the measurements in Paper I and
those in our work, respectively. In the {\it Right} panel, the symbols
are the same as those in the {\it Left} panel. We find that the
dispersion of the velocity becomes smaller when the GCs are locate
further from the center of the galaxy with larger projected radius
$r_p$. It can be seen that the dispersion of the radial velocity becomes 
smaller when the projected radius $r_p$ is larger.

Subsequently, all the spectra were shifted to the zero radial
velocity and smoothed to the wavelength dependent Lick resolution
with a variable-width Gaussian kernel following the definition of
\cite{wo97}, i.e. 11.5 {\AA} at 4000 {\AA}, 9.2 {\AA} at 4400 {\AA},
8.4 {\AA} at 4900 {\AA}, 8.4 {\AA} at 5400 {\AA}, 9.8 {\AA} at 6000
{\AA}. Indeed, we measured all the 25 types of Lick indices strictly
by using the parameters and formulae from \citet{w94a} and
\citet{wo97}. The uncertainty of each index was estimated based on
the analytic formulae (11)$-$(18) of \citet{car98}.

\begin{figure}
  \resizebox{\hsize}{!}{\rotatebox{-90}{\includegraphics{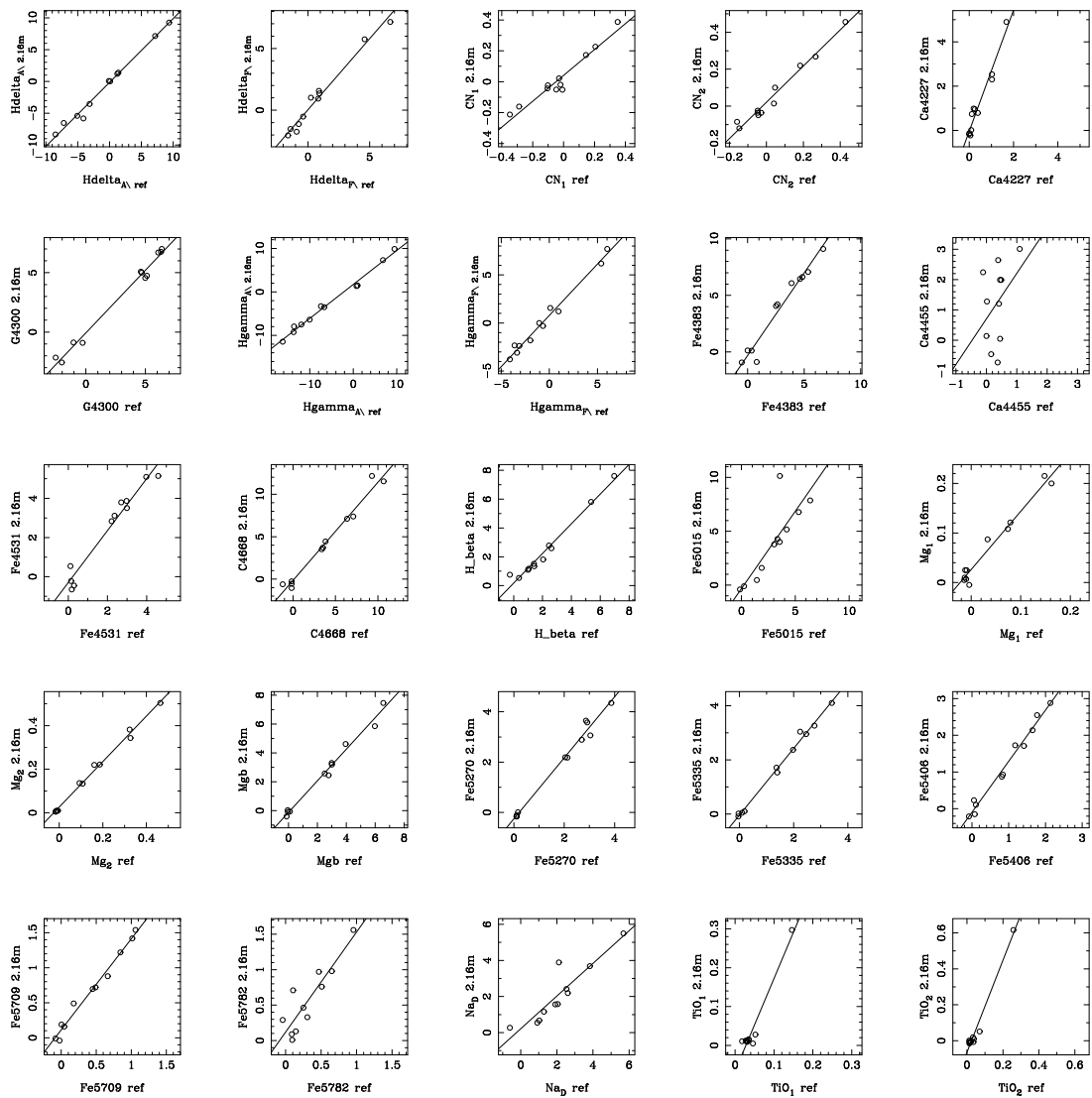}}}
  \caption{Calibrations of index measurements from the eleven standard stars
    of 2.16m raw spectra with those from reference \citet{wo97,w94a}.
    The linear fit coefficients of Eq.~\ref{eq3} have been derived to
    be used for calibrating our raw data to the Lick index system.}
  \label{fig4}
\end{figure}

Eq.~\ref{eq1} is the linear fit formula for calibrating the raw
measurements of our 2.16m data to the standard Lick index system.
The eleven standard stars are utilized for the fitting (Please see
Figure~\ref{fig4}) and the results are listed in Table~\ref{t3.tab}.
\begin{equation}
  \rm
  EW_{ref} = {\it a} + {\it b}  \cdot EW_{raw}
  \label{eq1}
\end{equation}

\begin{table}[ht!!!]
  \small
\centering
\begin{minipage}[]{100mm}
  \caption[]{The Linear Fit Coefficients $a$ and $b$ in Eq.~\ref{eq3} for transformations of the 2.16m data to the Lick index system.}\label{t3.tab}\end{minipage}
\tabcolsep 3mm
\begin{tabular}{lcc}
  \hline\noalign{\smallskip}
  Index & $a$ & $b$ \\
  \hline\noalign{\smallskip}
$\rm H\delta_A$ (\AA)& $-0.15\pm0.19$ & $1.00\pm0.04$ \\
$\rm H\delta_F$ (\AA)& $ 0.04\pm0.15$ & $1.15\pm0.06$ \\
$\rm  CN1$ (mag)     & $ 0.04\pm0.01$ & $0.84\pm0.07$ \\
$\rm  CN2$ (mag)     & $ 0.02\pm0.01$ & $0.98\pm0.05$ \\
$\rm Ca4227$ (\AA)   & $-0.04\pm0.14$ & $2.73\pm0.21$ \\
$\rm  G4300$ (\AA)   & $-0.06\pm0.19$ & $1.05\pm0.04$ \\
$\rm H\gamma_A$ (\AA)& $ 1.73\pm0.26$ & $0.78\pm0.03$ \\
$\rm H\gamma_F$ (\AA)& $ 0.79\pm0.16$ & $1.07\pm0.05$ \\
$\rm Fe4383$ (\AA)   & $-0.32\pm0.36$ & $1.46\pm0.10$ \\
$\rm Ca4455$ (\AA)   & $ 0.71\pm0.56$ & $1.50\pm1.21$ \\
$\rm Fe4531$ (\AA)   & $-0.30\pm0.24$ & $1.33\pm0.09$ \\
$\rm Fe4668$ (\AA)   & $-0.16\pm0.31$ & $1.16\pm0.06$ \\
$\rm  H\beta$(\AA)   & $ 0.17\pm0.16$ & $1.03\pm0.05$ \\
$\rm Fe5015$ (\AA)   & $-0.34\pm1.07$ & $1.44\pm0.30$ \\
$\rm  Mg1$ (mag)     & $ 0.03\pm0.01$ & $1.18\pm0.07$ \\
$\rm  Mg2$ (mag)     & $ 0.03\pm0.01$ & $1.04\pm0.03$ \\
${\rm  Mg}b$ (\AA)   & $-0.12\pm0.15$ & $1.09\pm0.04$ \\
$\rm Fe5270$ (\AA)   & $-0.25\pm0.11$ & $1.21\pm0.05$ \\
$\rm Fe5335$ (\AA)   & $-0.04\pm0.06$ & $1.23\pm0.03$ \\
$\rm Fe5406$ (\AA)   & $-0.10\pm0.08$ & $1.39\pm0.07$ \\
$\rm Fe5709$ (\AA)   & $ 0.11\pm0.03$ & $1.30\pm0.06$ \\
$\rm Fe5782$ (\AA)   & $ 0.12\pm0.10$ & $1.41\pm0.24$ \\
$\rm    NaD$ (\AA)   & $ 0.21\pm0.36$ & $0.91\pm0.14$ \\
$\rm TiO1$ (mag)     & $-0.07\pm0.01$ & $2.38\pm0.19$ \\
$\rm TiO2$ (mag)     & $-0.07\pm0.01$ & $2.56\pm0.14$ \\
  \noalign{\smallskip}\hline
\end{tabular}
\end{table}

\section{Fitting, analysis and results}
\label{fit.sec}

\subsection{Model description}

\citet{tmb} provided stellar population models including Lick
absorption line indices for various elemental-abundance ratios,
covering ages from 1 to 15 Gyr and metallicities from 1/200 to $3.5
\times$ solar abundance. These models are based on the standard
models of \citet{mar98}, with input stellar evolutionary tracks from
\citet{ccc97} and \citet{bo97} and a \citet{sal} stellar initial
mass function. \citet{tmk} improved the models by including
higher-order Balmer absorption-line indices. They found that these
Balmer indices are very sensitive to changes in the $\rm \alpha/Fe$
ratio for supersolar metallicities. The latest stellar population
model for Lick absorption-line indices \citep{tmj} is an improvement
on \citet{tmb} and \citet{tmk}. They were derived from the MILES
stellar library, which provides a higher spectral resolution
appropriate for MILES and SDSS spectroscopy, as well as flux
calibration. The models cover ages from 0.1 to 15 Gyr, $\rm [Z/H]$
from $-2.25$ to 0.67 dex, and $\rm [\alpha/Fe]$ from $-0.3$ to 0.5
dex. In our work, we fitted our absorption indices based on the
models of \citet{tmj}, by using the two sets of stellar evolutionary
tracks provided, i.e., \citet{ccc97} and Padova.

\subsection{Fits with stellar population models and the results}
\label{res.sec}

Similar to \citet{sha} and our Paper I, the $\chi^2-$minimization
routine was applied for fitting Lick indices with the SSP models to
derive the physical parameters. As we measured 25 different
types of Lick line indices listed in Table~\ref{t3.tab}, 
all indices were used for the fitting procedure. As \citet{tmj}
provide only 20 ages, 6 metallicity 
$\rm [Z/H]$, and 4 $\alpha$-element $\rm [\alpha/Fe]$ for the SSP
model, it is necessary to interpolate the original models to the
higher-resolution models for our needs. We performed the cubic
spline interpolations, using equal step lengths, to obtain a grid of
150 ages from 0.1 to 15 Gyr, 31 $\rm [Z/H]$ values from $-2.25$ to
0.67 dex, and 51 $\rm [\alpha/Fe]$ from $-0.3$ to $0.5$ dex,
which could make the fitted results smoother and more continuous.
Since \citet{w94b,gall09} pointed out the age-metallicity degenaracy
for most of the spectral feature indices measurements, which almost
remain the same when the percentage change ${\rm \Delta age/\Delta}
Z=3/2$. Therefore, it is necessary for us to constrain the metallicity
with the metal-sensitive indices before the fits.

Fortunately, \citet{gall09} provide two ways to measure the
metallicity from the metal-sensitive spectral indices directly. One
method is through combining the absorption line indices Mg and Fe,
$\rm [MgFe]$, which is defined as $\rm [MgFe] = \rm \sqrt{Mg{\it
    b}\cdot\langle Fe \rangle}$, where $\rm \langle Fe \rangle
=(Fe5270+Fe5335)/2$. Thus, 
the metallicity can be calculated from the formula below,
\begin{equation}
  \rm [Fe/H]_{[MgFe]}=-2.563+1.119[MgFe]-0.106[MgFe]^2\pm0.15.
  \label{eq2}
\end{equation}

The second way to obtain the metallicity from Mg2 is using a
polynomial in the following,
\begin{equation}
  \rm [Fe/H]_{Mg2}=-2.276+13.053Mg2-16.462Mg2^2\pm0.15.
  \label{eq3}
\end{equation}

Finally we obtained $\rm [Fe/H]_{avg}$ with uncertainty in
Table~\ref{t4.tab}, which is an average of the metallicities derived
from the metallicity Eqs.~\ref{eq2} and \ref{eq3}, respectively. The
averaged metallicity $\rm [Fe/H]_{avg}$ will be used to constrain
the metallicity in the fits to break the age-metallicity
trends/degeneracy. However, \citet{tmj} model only provide the    
metallicity parameters with $\rm [Z/H]$ and $\rm [\alpha/Fe]$,  thus
we need to find a relationship between the iron abundance$\rm
[Fe/H]$,  total metallicity $\rm [Z/H]$ and $\alpha$-element to iron
ratio $\rm [\alpha/Fe]$, which we can replace $\rm [Fe/H]$ with $\rm
[Z/H]$ and $\rm [\alpha/Fe]$ in the fit procedure. In fact,
\citet{tmb} give the relation in Eq.~\ref{eq4}.
 
\begin{table}
\bc
\begin{minipage}[]{100mm}
  \caption[]{The metallicities $\rm [Fe/H]$ derived from the spectral indices
  $\rm[MgFe]$, Mg2.}\label{t4.tab}
\end{minipage}
\small
\tabcolsep 1mm
\begin{tabular}{cc}
  \hline\noalign{\smallskip}
  Name &$\rm [Fe/H]_{avg}$  \\
  \hline\noalign{\smallskip}
  B289      &  $ -1.83\pm0.27$  \\
  B290      &  $ -0.56\pm0.63 $ \\
  H11       &  $ -0.49\pm0.58 $ \\
  H18       &  $ -1.35\pm0.65 $ \\
  SK108A &  $ -2.35\pm0.22 $ \\
  SK112A &  $ -1.62\pm0.43 $ \\
  MGC1    &  $ -2.06\pm0.33 $ \\
  H25       &  $ -2.74\pm0.47 $ \\
  \noalign{\smallskip}\hline
\end{tabular}
\ec \tablecomments{0.86\textwidth}{Here we define $\rm
[Fe/H]_{avg}=\frac{[Fe/H]_{[MgFe]} + [Fe/H]_{Mg2}}{2}$}
\end{table}

\begin{equation}
  \rm
[Z/H] = [Fe/H] + 0.94 [\alpha/Fe]
 \label{eq4}
\end{equation}

Here we would like to draw reader's attention that although the
metallicity $\rm [Fe/H]$ has been determined 
primarily, there are still many different ways to combine $\rm
[Z/H]$ and $\rm [\alpha/Fe]$ in the parameter grid of the
model. Therefore, we still need to fit the age, $\rm [Z/H]$ and $\rm
[\alpha/Fe]$ simultaneously. Here, we would like to constrain
the metallicity in the fits for $\rm |[Fe/H]_{fit} -[Fe/H]_{avg}|
\le 0.3$ dex, which is the typical metallicity uncertainty for the
observations and it will make the fits more reasonable.
Like the Paper I, the physical parameters ages, metallicities $\rm
[Z/H]$, and $\rm [\alpha/Fe]$ can be determined by comparing the
interpolated stellar population models with the observational
spectral feature indices by employing the $\chi^2-$minimization
method below,  
\begin{equation}
  \chi^2_{\rm min}={\rm
    min}\left[\sum_{i=1}^{25}\left({\frac{L_{\lambda_i}^{\rm
          obs}-L_{\lambda_i}^{\rm model}(\rm age,[Z/H],[\alpha/Fe])}
      {\sigma_i}}\right)^2\right],
\label{eq5}
\end{equation}
where $L_{\lambda_i}^{\rm model}(\rm age,[Z/H],[\alpha/Fe])$ is the
$i^{\rm th}$ Lick line index in the stellar population model for
age, metallicity $\rm [Z/H]$, and $\rm [\alpha/Fe]$, while
$L_{\lambda_i}^{\rm obs}$ represents the observed calibrated Lick
absorption-line indices from our measurements and the errors
estimated in our fitting are given as follows,
\begin{equation}
\sigma_i^{2}=\sigma_{{\rm obs},i}^{2}+\sigma_{{\rm model},i}^{2}.
\label{eq6}
\end{equation}
Here, $\sigma_{{\rm obs},i}$ is the observational uncertainty while
$\sigma_{{\rm model},i}$ is the uncertainty associated with the models of
\citet{tmj}. These two types of uncertainties have been both considered in
our fitting procedure.

Table~\ref{t5.tab} lists the fitted ages, $\rm [Z/H]$ and $\rm
[\alpha/Fe]$ with different evolutionary tracks of \citet{ccc97} and
Padova, respectively. In addition, we calculated the $\rm [Fe/H]_{cassisi}$ and
$\rm [Fe/H]_{padova}$ by applying the Eq.~\ref{eq4} to the fitted $\rm
[Z/H]$ and $\rm [\alpha/Fe]$. For the reason of keeping
consistency with Paper I, we adopted the metallicity $\rm
[Fe/H]_{cassisi}$ in the following statistics and analysis. From
Table \ref{t5.tab}, we found that the ages, $\rm 
[Z/H]$ and the $\alpha$-element $\rm [\alpha/Fe]$ fitted from
either \citet{ccc97} or Padova tracks are consistant with each other.
Besides, it is worth noting that all of our sample halo GCs are older
than 10 Gyr in both evolutionary tracks except B290 (5.5 to 5.8 Gyr),
which is older than 2 Gyr and it should be 
identified as the "old" in \citet{cw09}. Thus, it indicates that
these halo clusters formed at the early stage of the galaxy
formation, which agrees well with the previous findings.

\begin{table}
\bc
\begin{minipage}[]{100mm}
  \caption[]{The $\chi^2-$minimization Fitting Results Using \citet{tmj} Models
    with \citet{ccc97} and Padova Stellar Evolutionary Tracks, respectively.}\label{t5.tab}
\end{minipage}
\scriptsize
\tabcolsep 1mm
\begin{tabular}{ccccccccc}
  \hline\noalign{\smallskip}
  &\multicolumn{4}{c}{Cassisi} & \multicolumn{4}{c}{Padova} \\
  \hline
  Name &Age & $\rm [Z/H]$ & $\rm [\alpha/Fe] $ & $\rm [Fe/H]$ &
  Age&$\rm [Z/H]$ & $\rm [\alpha/Fe]$ & $\rm [Fe/H]$ \\
  & (Gyr) & (dex) & (dex) &(dex) & (Gyr) & (dex) & (dex) & (dex) \\
  \hline
  \noalign{\smallskip}

  B289  &  $ 10.75\pm4.15$ & $-1.67 \pm 0.23$& $0.34\pm0.16$  &
  $-2.09 \pm 0.27$& $11.70\pm2.80$ &  $-2.07\pm0.18$ & $-0.12\pm0.18$ 
  & $-2.13\pm0.25$\\  
  B290  &  $ 5.80\pm2.40 $ & $-0.99\pm0.05$ & $-0.26\pm0.05$ &
  $-0.85\pm0.07$ & $5.50\pm0.40$&  $-1.33\pm0.38$ & $-0.26\pm0.05$ &
  $-0.85\pm0.39$ \\ 
  H11   &  $ 13.75\pm1.25 $ & $ 0.09\pm0.32$ & $0.08\pm0.05$ & $
  -0.19\pm0.33$ & $ 13.60\pm0.20$ & $ -0.10\pm0.24$ & $0.00\pm0.06$ &
  $-0.21\pm0.24$\\ 
  H18   &  $ 13.45\pm1.45 $ & $-0.47\pm0.37$ & $0.48\pm0.02$ &
  $-1.07\pm0.37$ & $13.60\pm0.20$ & $-0.50\pm0.24$ & $0.48\pm0.02$ &
  $-1.07\pm0.24$\\ 
  SK108A&  $ 13.60\pm0.30 $ & $-1.53\pm0.18$ & $0.28\pm0.22$ &
  $-2.09\pm0.28$& $13.55\pm0.45$ & $-1.48\pm0.23$ &  $0.27\pm0.24$ &
  $-2.09\pm0.32$\\  
  SK112A&  $ 11.10\pm3.90 $ & $-1.33\pm0.38$ & $0.25\pm0.25$  &
  $-1.35\pm0.45$ &  $11.70\pm3.30$ & $-1.51\pm0.47$ &  $0.10\pm0.40$ &
  $-1.42\pm0.61$\\  
  MGC1  &  $ 13.30\pm0.80 $ & $-1.39\pm0.14$ & $0.42\pm0.08$ &
  $-1.76\pm0.16$&  $12.90\pm1.30$ & $ -1.39\pm0.14$ & $0.42\pm0.08$ &
  $-1.76\pm0.16$ \\ 
  H25   &  $ 13.60\pm0.30 $ & $-1.98\pm0.20$ & $0.50\pm0.00$  &
  $-2.45\pm0.20$ & $13.50\pm0.50$ & $ -2.03\pm0.05$ & $0.50\pm0.00$ &
  $-2.45\pm0.05$ \\ 
  \noalign{\smallskip}\hline
\end{tabular}
\ec
\end{table}

Actually, \citet{mac10} conclude that the metal abundance of MGC1 is
about $\rm [Fe/H]=-2.3$ and age is 12.5 to 12.7 Gyr through the
color-magnitude diagram fitting. The age estimated agree well with
our results while the metallicity is lower than our estimate $\rm
[Fe/H]_{avg}=-2.06\pm0.33$ in Table~\ref{t4.tab} or $\rm
[Fe/H]_{cassis}=-1.76\pm0.16$ in Table~\ref{t5.tab}. Nevertheless,
\citet{alv} found that the metallicity 
$\rm [Fe/H]=-1.37\pm0.15$ by combining the spectroscopic data and
the photometric data, which is higher than our estimate. Hence, it
can be seen that our result is just between the two results,
suggesting that our result agrees with the previous conclusions.

\subsection{Metallicity Properties of Outer Halo}
\label{md.sec}

The metallicity gradient of the halo star clusters and stars are
important to the formation and enrichment processes of their
host galaxy. Basically, there are two possible scenarios for the
galaxy formation. One is that the halo stars and clusters should
feature large-scale metallicity gradients if the enrichment timescale
is shorter than the collapse time, which may be due to the galaxy
formation as a consequence of a monolithic, dissipative, and rapid
collapse of a single massive, nearly spherical, spinning gas cloud
\citep{eggen62,bh00}. The other one is a chaotic scheme for early
galactic evolution, when the loosely bound
pre-enriched fragments merge with the protogalaxy during a very long
period of time, in which case a more homogeneous metallicity
distribution should develop \citep{sz78}. However, most galaxies are
believed to have formed through a combination of these scenarios.

\citet{van69,hsv82} showed that there is little or no evidence for a
general radial metallicity gradient for GCs within a radius of 50
arcmin. However, studies including \citet{hbk91,per02,fan08} support
the possible existence of a radial metallicity gradient for the
metal-poor M31 GCs, although the slope is not very significant.
\citet{per02} suggest that the gradients is $-0.017$ and $-0.015$
dex arcmin$^{-1}$ for the full sample and inner metal-poor clusters.
More recently, \citet{fan08} found that the slope is $-0.006$ and
$-0.007$ dex arcmin$^{-1}$ for the metal-poor subsample and whole
sample while the slope approaches zero for the metal-rich subsample.
Nevertheless, all these studies are based on GCs that are located
relatively close to the center of the galaxy, usually at projected
radii of less than 100 arcmin. Recently, \citet{h11} investigated
the metallicity gradient for 15 halo GCs to $r_{\rm p}=117$ kpc with
the metallicity derived from the CMD fittings
\citet{mac06,mac07,mac10} and the authors found that the metallicity
gradient becomes insignificant if one halo GC H14 is excluded in
their Figure~6. We found that our result is consistent well with the
previous findings of \citet{h11}. In Paper I, we found the slope of
metallicity gradient is $-0.018\pm0.001$ dex kpc $^{-1}$ for the
halo clusters sample extended to $r_{\rm p} \sim117$ kpc from the
galaxy center. Further, the slope turns to be $-0.010\pm0.002$ dex
kpc $^{-1}$ if only considering the clusters $r_p>25$ kpc.

Since we have spectroscopic observations of eight more halo confirmed
clusters, it is interesting to check if the metallicity
distribution/spatial gradient would change with
an enlarged halo clusters sample. For the new
observed data, as we recalled in \S \ref{res.sec}, only MGC1 have
the previous metallicity measurements from the literatures, which are
very different for different works and our measurement is just the
median value. Thus, we adopted our measurement. Finally we have
a metallicity sample of 391 entries in total.

\begin{figure}
\resizebox{\hsize}{!}{\rotatebox{0}{\includegraphics{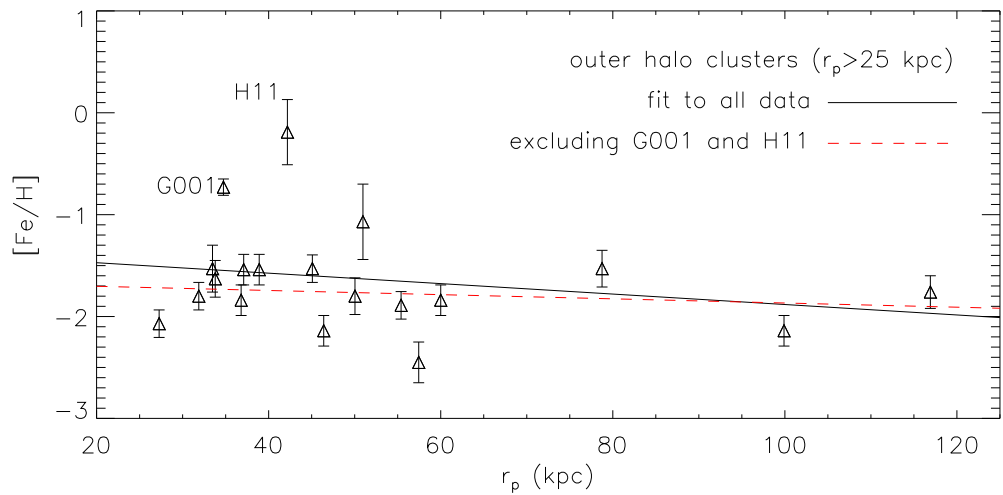}}}
\caption{Metallicities $\rm [Fe/H]$ versus projected radii for the outer
  halo GCs with $r_{\rm p} >25$ kpc from the center of the galaxy. The
  slope of the linear fitting is $-0.005\pm0.005$ dex kpc$^{-1}$
  (black solid line). However, if the two highest metallicity GCs G001
  and H11 are excluded, the slope turns out to be
  $-0.002\pm0.003$ dex kpc$^{-1}$ (red dashed line).}
\label{fig5}
\end{figure}

Figure~\ref{fig5} shows the metallicity as a function of projected
radius from the galaxy center for all outer GCs with spectroscopic
metallicity with $r_{\rm p} >25$ kpc from the galaxy center. The
slope of a linear fit is $-0.005\pm0.005$ dex kpc$^{-1}$, which is
marked with a solid black line. However, if the two highest
metallicity star clusters G001 and H11 are excluded, the slope turns
out to be $-0.002\pm0.003$ dex kpc$^{-1}$, which is shown with the
red dashed line. Thus, both of the cases suggest there is none
metallicity gradient for the M31 outer halo clusters when $r_{\rm p}
>25$ kpc, which agree with the conclusion of Paper I. Therefore it
seems that the ``fragments merging'' scenario dominated in the outer
halo during the galaxy formation stage.

It should be noted that the metallicity gradient is fitted based
on the data of our observations and the literature and the
metallicities from different literature may not the same.  For
instance, the metallicity of G001 is $\rm [Fe/H]=-1.08\pm0.09$ in
\citet{hbk91} while  $\rm [Fe/H]=-0.73\pm0.15$ in
\citet{gall09}. Thus we wonder how the slope would change when the
data is changed. We simulated ten sets of random data from 
$\sigma=-0.5$ to 0.5 and added them to the metallicities that we used 
in Figure~\ref{fig5} and then refit the slopes again for ten times 
separatedly and the results are shown in Table~\ref{t6.tab}. It shows  
that the slope dose not change significantly when the simulated errors were 
added, suggesting that the slope is stable even the data from different
measures.

\begin{table}
\bc
\begin{minipage}[]{100mm}
  \caption[]{The slopes of metallicity gradient by adding the random
    errors to the data.}\label{t6.tab}
\end{minipage}
\small
\tabcolsep 1mm
\begin{tabular}{ccc}
  \hline\noalign{\smallskip}
  No. &$k_{all}$  & $k_{<-1}$\\
  \hline\noalign{\smallskip}
  1    &  $ -0.013\pm0.010$  & $-0.013\pm0.011$ \\
  2    &  $ -0.003\pm0.010$  & $ 0.000\pm0.014$\\
  3    &  $ -0.011\pm0.012$  & $-0.008\pm0.012$\\
  4    &  $ -0.009\pm0.009$  & $ 0.000\pm0.012$\\
  5    &  $ -0.003\pm0.013$  & $ -0.036\pm0.021$\\
  6    &  $ -0.002\pm0.012$  & $ 0.004\pm0.022$\\
  7    &  $ -0.004\pm0.013$  & $ -0.002\pm0.022$\\
  8    &  $ -0.009\pm0.012$  & $ 0.009\pm0.015$\\
  9    &  $ -0.005\pm0.010$  & $ -0.008\pm0.017$\\
10    &  $ -0.013\pm0.011$  & $ 0.003\pm0.008$\\
  \noalign{\smallskip}\hline
\end{tabular}
\ec 
\end{table}

\begin{figure}
\resizebox{\hsize}{!}{\rotatebox{0}{\includegraphics{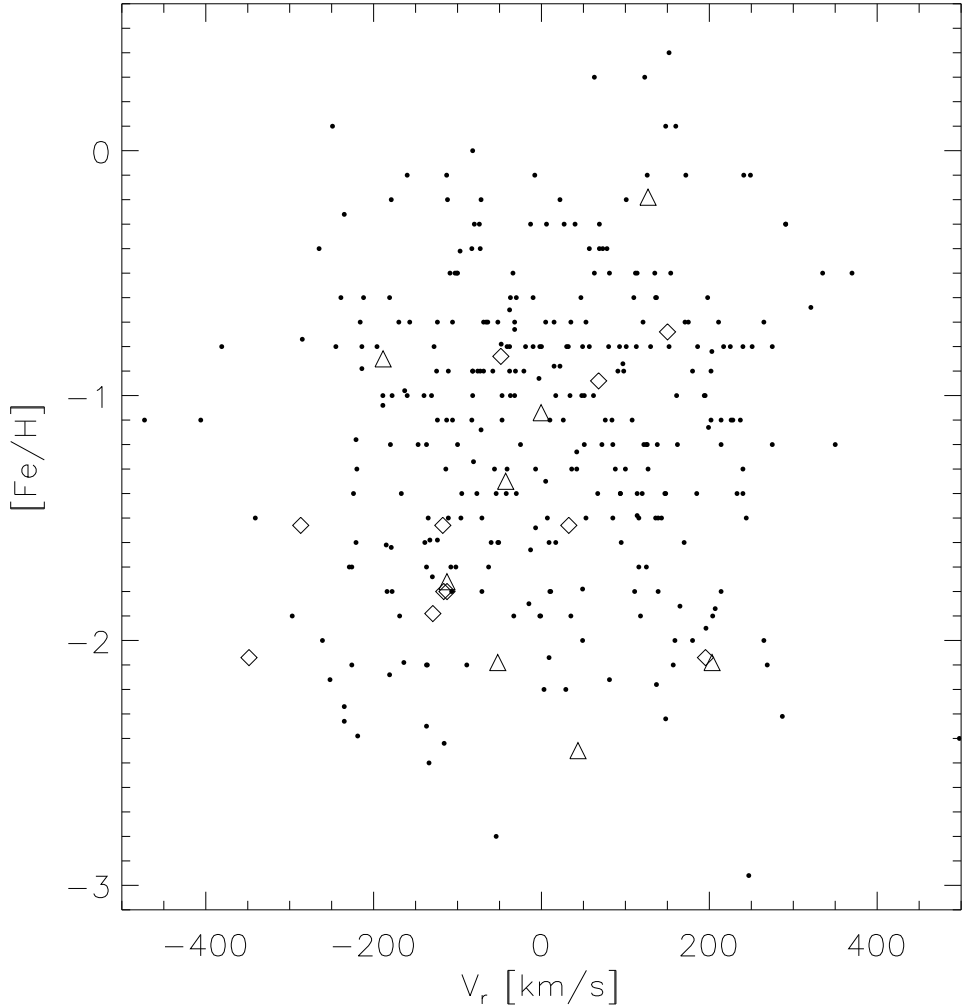}}}
\caption{Metallicity $\rm [Fe/H]$ versus radial velocity $V_r$ (corrected
for the systemic velocity of M31) for all the GCs with spectroscopic
metallicities and radial velocity. The small points are from the
literature; the squares are from Paper I; the triangles are from our
measurement.} \label{fig6}
\end{figure}

Figure~\ref{fig6} shows the relationship between the metallicities
and the radial velocities $V_r$ which have been corrected for the
systemic velocity of the M31 galaxy. The spectroscopic metallicities
are from the literature \citep{hbk91,bh00,per02,gall09,cw11}, Paper
I as well as this work and the radial velocities $V_r$ are from the
RBC v.4, Paper I and this work. It seems that there is no any
relationship between the metallicities versus the radial velocities
$V_r$.

\begin{figure}
\resizebox{\hsize}{!}{\rotatebox{0}{\includegraphics{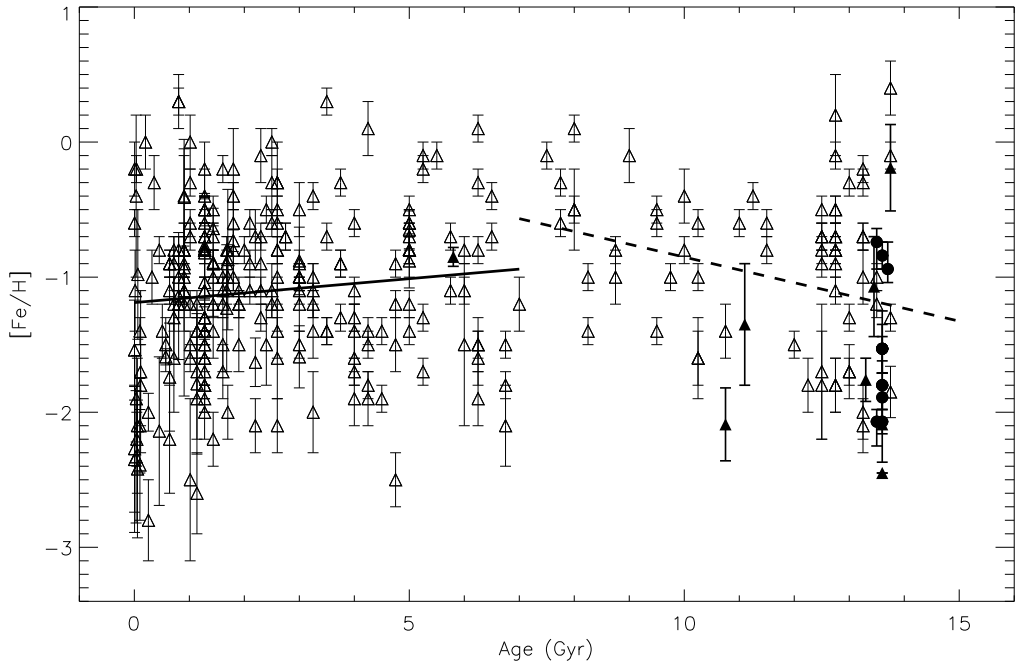}}}
\caption{Metallicity $\rm [Fe/H]$ versus ages for all the clusters with
  spectroscopic metallicity and age estimates. The open triangles are the data
  from the literature; the filled circles are the data from Paper I;
  the filled triangles are the data from this
  work. The solid line represents the linear fit of GCs younger than 7
  Gyr while the dashed line is the fit for the GCs older than 7 Gyr.}
\label{fig7}
\end{figure}

Figure~\ref{fig7} shows the metallicities versus ages of the GCs. The
metallicities are from the literature
\citep{hbk91,bh00,per02,gall09,cw11}, Paper I as well as this work
and the ages are from the \citet{fan10}, Paper I and this
work. We would like to see if there is any relationship between the
ages and metallicities for these GCs. Actually we find that the
relationships are different for the GC populations with different age. 
The slope of the GCs younger than 7 Gyr is $k=0.035\pm0.021$ while the
slope of the GCs older than 7 Gyr is $k=-0.095\pm0.034$, which is
$\sim3\sigma$ significant level. It suggests that for the GCs younger
than 7 Gyr, there is no relationship between the age and metallicity while
for the clusters older than 7 Gyr, it seems that the older GCs are
more metal-poor (lower metallicity) and the younger GCs are more
metal-rich (higher metallicity).

\begin{figure}
\resizebox{\hsize}{!}{\rotatebox{0}{\includegraphics{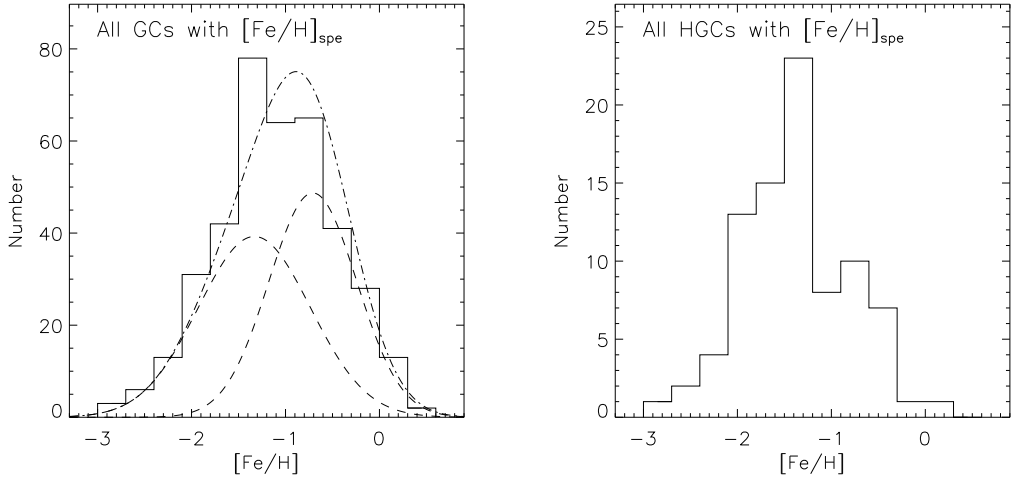}}}
\caption{Metallicity distributions with bin size of 0.3 dex. {\it Left}: 
  all the GCs with spectroscopic metallicities. The mixture-model KMM test 
  was applied to divide them to two groups. {\it Right}: all the HGCs with 
  spectroscopic metallicities.}
\label{fig8}
\end{figure}

Previously, many astronomers found the significant bimodal case in
the metallicity of M31 GC distribution by applying the mixture-model
KMM test \citep{ash94}. \citet{ash93,bh00,per02} found the
proportion of the metal-poor and metal-rich group is $\sim 2:1$ to
$\sim 3:1$ with the peak positions of $\rm [Fe/H]\approx-1.5$ and
$-0.6$, respectively. \citet{fan08} examined the bimodality of
metallicity distribution with a larger sample and the authors found
the proportion is $\sim 1.5:1$ and the the peak positions are $\rm
[Fe/H]\sim-1.7$ and $\sim-0.7$, respectively. However, the recent
work of \citet{cw11} suggests that there is no significant
bimodality or trimodality for metallicity distribution with a sample
of 322 M31 GCs, most of which have spectroscopic metallicity with
high S/N ratio. Since we have new observation data and a larger
spectroscopic data sample, we are able to reexamine the bimodality
of the metallicity distributions of M31 GCs. Figure~\ref{fig8} shows
the metallicity distributions of the GCs and the HGCs, respectively.
In the {\it Left} panel, the sample includes all the GCs which have
spectroscopic metallicity from the literature
\citep{hbk91,bh00,per02,gall09,cw11} and Paper I as well as this
work. In total, there are 386 GCs with spectroscopic metallicity in
the distribution. We applied the mixture-model KMM algorithm to the
dataset and it returns an insignificant bimodality with {\it
p}-value $=0.369$, which means that a bimodal distribution is
preferred over a unimodal one at 63.1\% confidence level. The
numbers of the metal-poor group and the metal-rich group are
$N1=196$, $N2=190$, respectively and the mean values of the two
groups are $\rm [Fe/H]_1=-1.43$ ($\sigma_1^2=0.327$) and $\rm
[Fe/H]_2=-0.73$ ($\sigma_2^2=0.215$), respectively. As we can see
from the plot, the proportion of the metal-poor and metal-rich group
is $\sim 1:1$, which is lower than the published results. The reason
why the bimodal case becomes more insignificant with larger sample
may be that more intermediate metallicity GCs, which is between the
two metallicity peaks, have been discovered and those intermediate
metallicity GCs cause the distribution to be unlikely a bimodal or
trimodal distribution. Therefore, the previous works found that the
metallicity distributions of M31 GCs is like that of the Milky Way
and more recent works with more data show that they are less similar
to each other, which may indicate that the formations of the two GC
system was substantially different. In the {\it Right} panel, it
show the metallicity distribution of the HGCs and obviously the
metal-poor GCs dominate in the distribution.

\begin{figure}
\resizebox{\hsize}{!}{\rotatebox{0}{\includegraphics{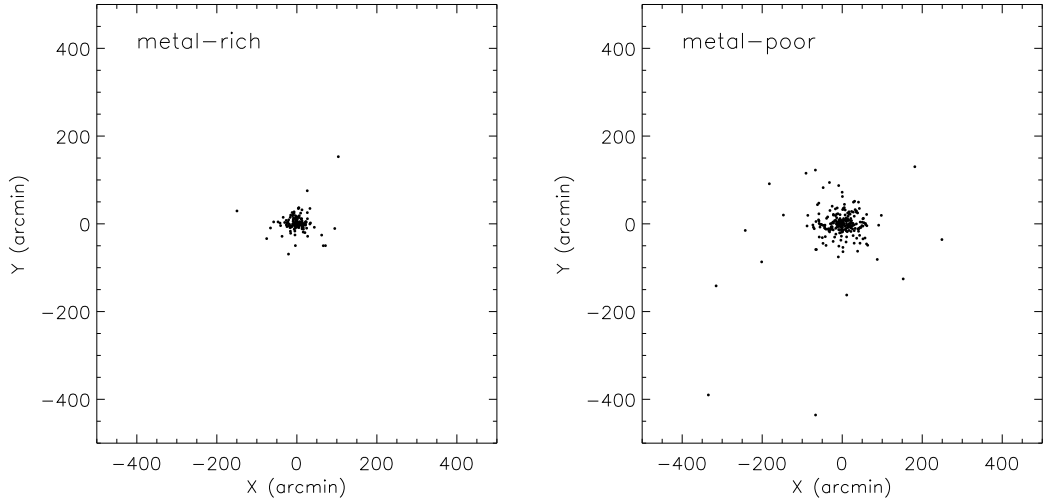}}}
\caption{The spatial distributions of HGCs with different
  metallicities. {\it Left}: metal-rich GCs; {\it Right}:
  metal-poor GCs. The two groups were divided by the KMM test of
  Figure~\ref{fig8}.}
\label{fig9}
\end{figure}

As the M31 GCs have been divided into two different groups by the
KMM test in the metallicity distribution of Figure~\ref{fig8}, we
would like to examine the spatial distributions of the two groups
with different metallicity. Figure~\ref{fig9} plots the spatial
distributions of the metal-rich and metal-poor groups. Note that the
metal-poor group appear to occupy a more extended halo and much more
widely spatially distributed while the metal-rich group is more
centrally concentrated, which is consistent with the conclusions of
\citet{per02,fan08}.

\begin{figure}
\resizebox{\hsize}{!}{\rotatebox{0}{\includegraphics{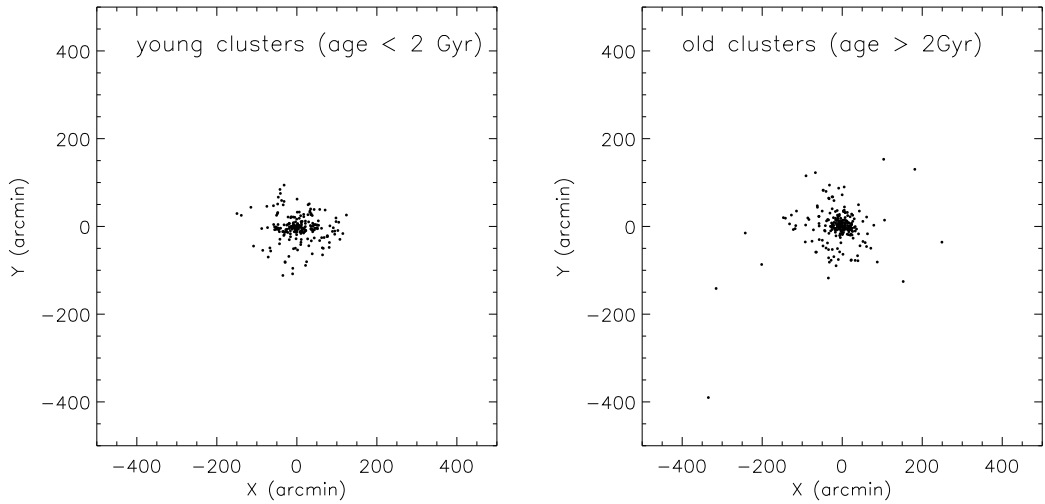}}}
\caption{The spatial distributions of HGCs with young and old
  populations, respectively. {\it
    Left}: young clusters with age $<2$ Gyr; {\it Right}: old clusters
  with age $>$ 2 Gyr.}
\label{fig10}
\end{figure}

Since we have the age estimates of the halo GCs in M31, we are
curious about whether the spatial distributions of the young and old
populations are the same or not. Here we used the definition of "old
population" for age $>2$ Gyr and the "young population" for age $<2$
Gyr as that did in \citet{cw09}. For the purpose of enlarging our
sample, the age estimates for M31 GCs in \citet{fan10} and Paper I
are also merged into our sample. Figure~\ref{fig10} plots the young
and old population spatial distributions, respectively. It is
obvious that the young population is more centrally concentrated and
it traces the disk shape of the galaxy well. However, the spatial
distribution of the old population is more dispersive and it seems
that they do not trace the disk shape of the galaxy.

\section{Summary and Conclusions}
\label{sum.sec}

This is the second paper of our serial works for M31 halo globular 
clusters. In Paper I, we mainly focus on the fits method and the metallicity 
gradient for the M31 GC sample. In this paper, we focus on the dynamics, 
metallicity and age, and their distributions as well as the relationships 
between these parameters. 

We selected eight more confirmed and bright GCs in the halo of M31 from 
RBC v.4 and observed them with the OMR spectrograph on 2.16 m telescope at 
Xinglong site of NAOC in the fall of 2011. These star clusters are located 
in the halo of galaxy at a projected radius of $\sim14$ to $\sim117$ kpc 
from the galactic center, where the sky background is dark so that they 
can be observed in high signal-to-noise ratio.

For all our sample clusters, we measured all 25 Lick absorption-line indices
\citep[see the definitions in,][]{w94a,wo97} and fitted the radial velocities.
We found that distributions of the confirmed GCs and the halo GCs are basically 
symmetric to the systematic velocity of the galalxy.

Similar to \citet{sha} and our Paper I, we applied the $\chi^2-$minimization 
method to fit the Lick absorption line indices with the updated \citet{tmj} 
stellar population model in two stellar evolutionary tracks of Cassisi
and Padova, separately. The fitting results show that most of 
our sample clusters are older than 10 Gyr except B290$\sim 5.5$ Gyr 
and most of them are metal-poor with metallicity $\rm [Fe/H]<-1$ dex 
except H11 and H18, suggesting that these halo star clusters were born at 
the early stage of the galaxy's formation

Again, we would like to study the metallicity gradient of the halo GCs 
by merging more spectroscopic metallicity from our work, Paper I and 
the literature. We only considered outer halo clusters 
with $r_{\rm p}>25$ kpc and the fitted slope is $-0.005\pm0.005$ dex kpc$^{-1}$.
However, if two metal-rich outlier clusters G001 and H11 are excluded,
the slope is $-0.002\pm0.003$ dex kpc$^{-1}$, which does not change 
significantly. Furthermore, in order to eliminate the effect the errors of 
different observations, we added the random errors from $\sigma=-0.5$ to $0.5$ 
to the data and refit the slope agian for ten times. The result shows that the 
simulated errors do not affact the slope much. Thus it seems
that metallicity gradient for M31 outer halo clusters dose not exist, 
which agrees well with the previous findings \citep{h11} and Paper I.
This result may imply that the ``fragments merging'' scenario is dominated 
in the outer halo of the galaxy beyond 25 kpc from the center during the early 
stage of the galaxy formation. 

We do not find a relationship between metallicity and the radial velocity for 
M31 GCs sample. It seems that the metallicity is not a function of age for 
the GCs with age $<7$ Gyr while for the old GCs with age $>7$ Gyr there 
seems to be a trend that the older ones have lower metallicity. This conclusion 
is similar to that of \citet{fan06}, who found a possible general trend of the 
age-metallicity relation with a large scatter. In addition, we plot 
metallicity distributions with the largest sample of M31 GCs so far and 
it shows the bimodality is not significant compared to the previous work. 
This is also found by \citet{cw11}, who used the newly observed spectroscopic 
data. We also find that the number of the metal-poor and metal-rich groups 
becomes comparable while the previous works show that the number of metal-poor 
group is more than that of the metal-rich one. This may be due to many 
intermedate metallicty metallicity of \citet{cw11} have been merged into our 
sample for our statistics. The spatial distributions shows 
that the metal-rich group is more centrally concentrated while the 
metal-poor group is occupy a more extended halo and the young population 
is centrally concentrated while the old populaiton is more extended 
spatially to the outer halo. This is easy to be understood as the old GCs are 
usually metal-poor especially for the halo GCs of M31.

\normalem
\begin{acknowledgements}
  We are indebted to an anonymous referee for his/her thoughtfull comments 
  and insightful suggestions that improved this paper greatly.
  The authors are also grateful to the kind staff at the Xinglong 2.16m
  telescope for the support during the observations.
  This research was supported by National Natural Science
  Foundation of China through grants Nos. 11003021, 11073027 and 11073032.
\end{acknowledgements}

\appendix                  

\label{lastpage}

\end{document}